\documentclass[showpacs,preprintnumbers,amsmath,amssymb]{revtex4}

\usepackage{graphicx}% Include figure files
\usepackage{dcolumn}% Align table columns on decimal point
\usepackage{bm}% bold math
\usepackage{epsf}

\newcommand{\pfrac}[2]{\left(\frac{#1}{#2}\right)}
\def\eps{\epsilon}
\def\vareps{\varepsilon}

\begin{document}

\title{Note on the Normalization of Predicted GRB Neutrino Flux }

\author{Zhuo Li}
\affiliation{ Department of Astronomy/Kavli Institute for Astronomy
and Astrophysics,\\ Peking University, Beijing 100871, China \\
Key Laboratory for the Structure and Evolution of Celestial Objects,
Chinese Academy of Sciences, Kunming 650011, China}

%\date{\today}

\begin{abstract}
We note that the theoretical prediction of neutrinos from gamma-ray
bursts (GRBs) by IceCube overestimates the GRB neutrino flux,
because they ignore both the energy dependence of the fraction of
proton energy transferred to charged pions and the radiative energy
loss of secondary pions and muons when calculating the normalization
of the neutrino flux. After correction for these facts the GRB
neutrino flux is reduced, e.g., by a factor $\sim5$ for typical GRB
spectral parameter, and may be consistent with the present zero
event detected by IceCube. More observations are important to push
the sensitivity below the prediction and test whether GRBs are the
sources of ultra-high energy cosmic rays.
\end{abstract}

%Use showkeys class option if keyword
                              %display desired
\maketitle

IceCube has become the most sensitive TeV-scale neutrino telescope
that may reach the predicted neutrino flux from gamma-ray bursts
(GRBs). The continued non-detection in its 22 (IC22, \citep{ic22}),
40 (IC40, \citep{ic40}) and 59 (IC59 \citep{ic59}) string
configuration puts more and more stringent limits on GRB neutrino
flux. The IC40 limit is comparable to the theoretical prediction
\citep{ic40}, whereas the combined IC40 and IC59 limit is only 0.22
times the prediction \citep{ic59}. These limits start to put
interesting constraints on the GRB neutrino models of Waxman and
Bahcall \cite{wb97,wb99} and Guetta et al. (2004) \citep{Guetta04},
and challenges GRBs as the sources of ultra-high energy cosmic rays
(UHECRs). However, in this note we show that the approach that
IceCube \citep{ic22,ic40,ic59} takes in theoretical prediction is
somewhat different from that of Waxman and Bahcall and Guetta et al.
\citep{wb97,wb99,Guetta04}, leading to overestimate of GRB neutrino
flux by a factor of $\sim5$ for typical GRB parameters. This is
because of ignoring the effects of the energy dependence of charged
pion production and secondary pion/muon cooling on the normalization
of neutrino flux.

The approach taken by IceCube papers \citep{ic22,ic40,ic59} is
presented in the appendix of the IC22 paper \citep{ic22}. In their
approach, the muon neutrino flux from a GRB, ${\mathcal F}_\nu$,
(with neutrino oscillation the electron, muon and tau neutrinos
roughly share equal energy \cite{particledata}) is scaled to the
proton flux in the GRB, ${\mathcal F}_p$, as
\begin{equation}\label{eq:ic}
  {\mathcal F}_\nu^{\rm IC}/{\mathcal F}_p=\frac18f_{\pi,b}
\end{equation}
(this comes from eq [A8] of \cite{ic22}). Here $f_{\pi,b}\equiv
f_\pi(E=E_b)$ is the fraction of energy of protons with $E_b$
carried by charged pions, and $E_b$ is the energy of protons that
interact with photons with spectral-break energy $\eps_b$ at
$\Delta$ resonance \citep{wb97},
\begin{equation}
  E_b=1.3\times10^{16}\Gamma^2_{2.5}\eps^{-1}_{b,\rm MeV}\rm eV,
\end{equation}
where $\Gamma=10^{2.5}\Gamma_{2.5}$ is the bulk Lorentz factor of
the GRB, and $\eps_b=1\eps_{b,\rm MeV}$~MeV. The $f_{\pi,b}$ value
depends on the GRB properties, and varies from burst to burst (see
eqs [A7] and [A8] of \cite{ic22}). The factor $1/8$ is due to the
facts that one half of the $p\gamma$ interactions produce charged
pions, and that each generated neutrino is assumed to carry one
fourth of the secondary pion energy. The proton flux can be
normalized to gamma-ray flux by ${\mathcal F}_p=(1/f_e){\mathcal
F}_\gamma$ where $f_e$ is the ratio of accelerated proton to
electron energy. In what follows we show that the approximation in
eq (\ref{eq:ic}) is different from the models of Waxman and Bahcall
\cite{wb97,wb99} and Guetta et al \cite{Guetta04}, and leads to
overestimate of the neutrino flux.

For a flat proton distribution with index $p\approx2$
($E^2dn_p/dE\propto E^{2-p}$), and a typical GRB spectrum with
Band-function parameters, $\alpha_\gamma=1$ and $\beta_\gamma=2$,
$f_\pi(E)=f_{\pi,b}$ is valid for protons with $E>E_b$. However, for
$E<E_b$, $f_\pi\propto E$ reduces with $E$ decreasing because fewer
target photons at high energy \cite{wb97}. Thus by using
$f_\pi(E)=f_{\pi,b}$ in energies $E<E_b$ eq. (\ref{eq:ic})
overestimates the neutrino flux.

Moreover,  eq. (\ref{eq:ic}) also ignores the suppression of
neutrino production at high energies due to the radiative cooling of
secondary pions/muons\footnote{It is possible that the secondaries
get accelerated before cooling. However the Fermi shock acceleration
requires the (far)downstream secondaries catch up with and cross the
(mildly) relativistic shock, which may be of low probability and is
ignored here.}. The synchrotron cooling timescale is shorter than
the secondary decay time at energies above the cooling energy
\cite{wb99,Guetta04,ic22},
\begin{equation}
  E_{c,\pi/\mu}=2\times10^{17}(\eps_e/\eps_B)^{1/2}\Gamma^4_{2.5}\Delta
  t_{-2}L_{52}^{-1/2}\times\left\{\begin{array}{ll}
    1 & (\mu^{\pm})\\
    10 & (\pi^{\pm})
  \end{array}\right.\rm eV.
\end{equation}
Here $L=10^{52}L_{52}\rm erg\,s^{-1}$ is the GRB (isotropic)
luminosity, $\Delta t=10^{-2}\Delta t_{-2}$s is the GRB
variability time, and $\eps_e$ and $\eps_B$ are the fractions of
internal energy carried by postshock electrons and magnetic field,
respectively.

Thus the neutrino production is mainly contributed by protons with
$E_b<E<E_c$, which is only a fraction of the total accelerated
protons in energy. The distribution of the accelerated protons is
expected to be a power law between the minimum and maximum energy.
For mildly-relativistic GRB internal shocks, the minimum
accelerated proton energy might be
\begin{equation}
  E_{\min}\simeq\Gamma m_pc^2=3\times10^{11}\Gamma_{300}\rm eV.
\end{equation}
The maximum proton energy is determined by the limit of
synchrotron cooling for typical GRB parameters \cite{waxman95},
\begin{equation}
  E_{\max}=2.5\times10^{20}\Gamma_{300}^{5/2}\Delta t_{-2}^{1/2}\epsilon_e^{1/4}\epsilon_B^{-1/4}g^{-1/2}L_{52}^{-1/4}\rm
  eV,
\end{equation}
where $g\gtrsim1$ accounts for the uncertainty in particle
acceleration time.

Since the neutrino production is mainly contributed by protons with
$E_b<E<E_c$ where $f_\pi(E)=f_{\pi,b}$, the muon neutrino flux
${\mathcal F}_\nu$ is estimated to be
\begin{equation}
  {\mathcal F}_\nu\approx\int_{E_b}^{E_c}\frac18f_{\pi,b}E\frac{dn_p}{dE}dE
  =\frac{{\mathcal F}_\nu^{\rm IC}}{{\mathcal F}_p}\int_{E_b}^{E_c}E\frac{dn_p}{dE}dE
  \approx {\mathcal F}_\nu^{\rm
  IC}\frac{\ln(E_c/E_b)}{\ln(E_{\max}/E_{\min})},
\end{equation}
where ${\mathcal F}_p=\int_{E_{\min}}^{E_{\max}}E\frac{dn_p}{dE}dE$
and Eq (\ref{eq:ic}) have been used, and the last equation holds for
$p\approx2$. Given $E_{\max}/E_{\min}\sim10^9$ and
$E_{c,\pi}/E_b\sim10^2$, the correction to Eq (\ref{eq:ic}) is a
factor of ${\mathcal F}_\nu/{\mathcal F}_\nu^{\rm IC}\sim0.22$.

Below is more detailed calculation about ${\mathcal F}_\nu/{\mathcal
F}_\nu^{\rm IC}$. For a flat proton distribution, $p=2$, and a GRB
spectrum with Band-function parameters, $\alpha_\gamma$,
$\beta_\gamma$ and $\eps_b$, the source neutrino spectrum (before
neutrino oscillation) is \cite{wb97,wb99,Guetta04,ic22}, for muon
neutrinos from secondary pion decay,
\begin{equation}\label{eq:spec1}
  \frac{dn^\pi_{\nu_\mu}}{d\vareps}=n_0\times\left\{\begin{array}{ll}
    \pfrac{\vareps}{\vareps_b}^{-\alpha_\nu} & \vareps<\vareps_b\\
    \pfrac{\vareps}{\vareps_b}^{-\beta_\nu} &
    \vareps_b<\vareps<\vareps_{c,\pi}\\
    \pfrac{\vareps_{c,\pi}}{\vareps_b}^{-\beta_\nu}\pfrac{\vareps}{\vareps_{c,\pi}}^{-(\beta_\nu+2)} &
    \vareps>\vareps_{c,\pi}
  \end{array}\right.;
\end{equation}
for electron and muon neutrinos from secondary muon decay,
\begin{equation}\label{eq:spec2}
  \frac{dn^\mu_{\nu_e}}{d\vareps}=\frac{dn^\mu_{\nu_{\mu}}}{d\vareps}=n_0\\
  \times\left\{\begin{array}{ll}
    \pfrac{\vareps}{\vareps_b}^{-\alpha_\nu} & \vareps<\vareps_b\\
    \pfrac{\vareps}{\vareps_b}^{-\beta_\nu} &
    \vareps_b<\vareps<\vareps_{c,\mu}\\
    \pfrac{\vareps_{c,\mu}}{\vareps_b}^{-\beta_\nu}\pfrac{\vareps}{\vareps_{c,\mu}}^{-(\beta_\nu+2)} &
    \vareps_{c,\mu}<\vareps<\vareps_{c,\pi}\\
    \pfrac{\vareps_{c,\mu}}{\vareps_b}^{-\beta_\nu}\pfrac{\vareps_{c,\pi}}{\vareps_{c,\mu}}^{-(\beta_\nu+2)}\pfrac{\vareps}{\vareps_{c,\pi}}^{-(\beta_\nu+4)} &
    \vareps>\vareps_{c,\pi}
  \end{array}\right..
\end{equation}
There is no tau neutrino $\nu_\tau$ generated in $p\gamma$
interactions, and the small effect of kaon production on neutrino
flux is neglected. Here the power law indices are
$\alpha_\nu=3-\beta_\gamma$ and $\beta_\nu=3-\alpha_\gamma$, and the
normalization is
\begin{equation}
  n_0\equiv\frac{dn_\nu(\vareps=\vareps_b)}{d\vareps}=50f_{\pi,b}\frac{dn_p(E=E_b)}{dE}.
\end{equation}
Note the factor $50=\frac12\times\frac1{0.2}/0.05$ is resulted
from the facts that (i) $1/2$ of $p\gamma$ interactions produce
charged pions; (ii) since the pion carries 0.2 of the proton
energy, the muon neutrino number (after oscillation) per proton is
$f_{\pi,b}/0.2$; and (iii) a single neutrino carries
$0.2\times\frac14=0.05$ of the proton energy, $\vareps_b=0.05E_b$.
Similarly, we assume the other break energies in the neutrino
spectrum, $\vareps_{c,\pi}\approx0.05E_{c,\pi}$, and
$\vareps_{c,\mu}\approx0.05E_{c,\mu}$.

Considering neutrino oscillation, the muon neutrino spectrum
(including $\nu_\mu$ and $\bar{\nu}_\mu$) detected on the Earth is
approximated as \cite{particledata}
\begin{equation}
  \frac{dn^{\oplus}_{\nu_\mu}}{d\vareps}\approx0.2\frac{dn_{\nu_e}}{d\vareps}
  +0.4\frac{dn_{\nu_\mu}}{d\vareps}+0.4\frac{dn_{\nu_\tau}}{d\vareps}
  =0.2\frac{dn^\mu_{\nu_e}}{d\vareps}
  +0.4\left(\frac{dn^\pi_{\nu_\mu}}{d\vareps}+\frac{dn^\mu_{\nu_\mu}}{d\vareps}\right).
\end{equation}
Note here $dn_{\nu_\tau}/d\vareps=0$ since no tau neutrino produced.
The muon neutrino flux is calculated as
\begin{equation}\label{eq:correctf}
  {\mathcal F}_\nu/{\mathcal F}_p=\frac{\int_{0.05E_{\min}}^{0.05E_{\max}}\vareps (dn^{\oplus}_{\nu_\mu}/d\vareps)d\vareps}{\int_{E_{\min}}^{E_{\max}}
  E(dn_p/dE)dE}.
\end{equation}
For various values of GRB spectral parameters, $\alpha_\gamma$,
$\beta_\gamma$ and $\eps_b$, we calculate the value of ${\mathcal
F}_\nu/{\mathcal F}_p$ with eqs. (\ref{eq:spec1}-\ref{eq:correctf}),
and show the ${\mathcal F}_\nu/{\mathcal F}_\nu^{\rm IC}$ value in
Table \ref{tab}. We find that the approximation of eq (\ref{eq:ic})
overestimates GRB neutrino flux by a few. Typically for GRBs with
$\alpha_\gamma=1$, $\beta_\gamma=2$, and $\eps^{\rm
obs}_b=\eps_b/(1+z)=0.2$~MeV, the correction factor is ${\mathcal
F}_\nu/{\mathcal F}_\nu^{\rm IC}=0.2$, i.e., the neutrino flux is
overestimated by a factor of 5 in \cite{ic22,ic40,ic59}.

Note the neutrino spectrum used here (eqs. \ref{eq:spec1} and
\ref{eq:spec2}) is different from the one by IceCube papers
\cite{ic22,ic40,ic59} at high energy; they assume a simple
steepening $dn^{\rm
ob}_{\nu_\mu}/d\vareps\propto\vareps^{-(\beta_\nu+2)}$ at
$\vareps>\vareps_{c,\mu}$ (see eq. [A3] in IC22 paper
\cite{ic22}), which underestimate the neutrino emission from pion
decay. Using their spectral shape, the neutrino flux is even
smaller than using ours, thus the correction factor is even
smaller.

\begin{table*}[h]
\centering
\begin{center}
\caption{The value of ${\mathcal F}_\nu/{\mathcal F}_\nu^{\rm IC}$
for various spectral parameter values. }
\begin{tabular}{cccccccc}
%\hline
% Parameters & & & & & & &  \\
\hline
$\alpha_\gamma$ & 1 & 0.5 & 1.5 & 1 & 1 & 1 & 1 \\
$\beta_\gamma$ & 2 & 2 & 2 & 1.5 & 3 & 2 & 2\\
$\eps_b^{\rm obs}/{\rm MeV}$ & 0.2 & 0.2 & 0.2 & 0.2 & 0.2 & 0.05 & 2\\
\hline
${\mathcal F}_\nu/{\mathcal F}_\nu^{\rm IC}$ & 0.20 & 0.12  & 0.56 & 0.25 & 0.18 & 0.14 & 0.30 \\
\hline
\end{tabular}\label{tab}
\end{center}
%\par
Note--The other parameters are $\Gamma=300$, $\Delta t=10^{-2}$s,
$L=10^{52}\rm erg\,s^{-1}$, $z=1.5$ and $E_{\max}=10^{21}$eV. The
redshift $z$ only has effect on the GRB spectral break,
$\eps_b=\eps_b^{\rm obs}(1+z)$.
\end{table*}

In summary, for a flat energy distribution of accelerated protons in
GRBs, $p\approx2$, the predicted neutrino flux using eq
(\ref{eq:ic}), so as the approach taken by \cite{ic22,ic40,ic59}, is
overestimated by a factor of ${\mathcal F}_\nu^{\rm IC}/{\mathcal
F}_\nu$, typically $\sim5$, compared to the models by
\cite{wb97,wb99,Guetta04}. So the IC40 limit of neutrino flux
\cite{ic40} actually should be above the prediction after
correction, while the combined IC40 and IC59 limit, which is claimed
to be 0.22 times of the prediction \cite{ic59}, could be consistent
with the correct prediction.

Some comments should be made here. In this brief note we point out
that when trying to test the models of \cite{wb97,wb99,Guetta04}
with data, the IceCube papers \cite{ic22,ic40,ic59} actually take an
approach different from the former models in calculating the
neutrino flux, leading to overestimate. However, the models of
\cite{wb97,wb99,Guetta04} consider only $\Delta$ resonance and
neglect the multi-pion production \cite{murase06,winter11a}, the
kaon production \cite{kaon} and the possible secondary particle
acceleration \cite{2ndacc}. All these facts may increase the
neutrino flux and somewhat compensate the overestimate. It is also
worth mentioning that if UHECRs are produced by the decay of
neutrons that escape from the GRB outflow, it is straightforward to
relate the neutrino flux with the observed UHECR flux, and avoid the
uncertainties in parameters, e.g., $f_\pi$ and $f_e$
\cite{n-escape}. If the assumption that only neutrons escape is
true, the neutrino flux will also increase.

Finally it should be stressed that there are large uncertainties in
the GRB neutrino models, which may cause lower $f_\pi$ and hence
lower neutrino flux, e.g., larger emission size or larger bulk
Lorentz factor. More observations by IceCube are required to push
the sensitivity below the uncertainties \cite{winter11b}, and test
the assumption of GRBs as the UHECR sources.

%Note, a newly posted paper \cite{winter11c} even found higher
%overestimate by IceCube, mainly due to the additional fact that
%energy of all photons are approximated by $\eps_b$ in
%\cite{Guetta04}, as followed by IceCube papers
%\cite{ic22,ic40,ic59}.

The author thanks the anonymous referees for valuable comments. This
work is partly supported by the Foundation for the Authors of
National Excellent Doctoral Dissertations of China and the Open
Research Program of Key Laboratory for the Structure and Evolution
of Celestial Objects, Chinese Academy of Sciences.

\end{document}